\documentclass[twocolumn,prl,amsmath,amssymb,showpacs,nosuperscriptaddress]{revtex4-1}

\usepackage{graphicx}
\usepackage{epstopdf}
\usepackage{hyperref}
\usepackage{url}
\usepackage{float}
\pdfoutput=1
\begin{document}
\newcommand{\ben}{\begin{equation}}
\newcommand{\een}{\end{equation}}

\title{Domain Wall in a Quantum Anomalous Hall Insulator as a Magnetoelectric Piston}

\author{Pramey Upadhyaya}  
\affiliation{Department of Physics and Astronomy, University of California, Los Angeles, California 90095, USA}

\author{Yaroslav Tserkovnyak}
\affiliation{Department of Physics and Astronomy, University of California, Los Angeles, California 90095, USA}

\begin{abstract}
We theoretically study the magnetoelectric coupling in a quantum anomalous Hall insulator state induced by interfacing a dynamic magnetization texture to a topological insulator. In particular, we propose that the quantum anomalous Hall insulator with a magnetic configuration of a domain wall, when contacted by electrical reservoirs, acts as a magnetoelectric piston. A moving domain wall pumps charge current between electrical leads in a closed circuit, while applying an electrical bias induces reciprocal domain-wall motion. This piston-like action is enabled by a finite reflection of charge carriers via chiral modes imprinted by the domain wall. Moreover, we find that, when compared with the recently discovered spin-orbit torque-induced domain-wall motion in heavy metals, the reflection coefficient plays the role of an effective spin-Hall angle governing the efficiency of the proposed electrical control of domain walls. Quantitatively, this effective spin-Hall angle is found to approach a universal value of $2$, providing an efficient scheme to reconfigure the domain-wall chiral interconnects for possible memory and logic applications. 

\end{abstract}


\maketitle

\textit{Introduction}.|Motivated by the possibility of realizing unique magnetoelectric phenomena, breaking of time-reversal symmetry in recently discovered topological insulators (TIs) \cite{Moore2010, *TI_rev_east, *TI_rev_west} has become one of the central themes of theoretical and experimental  research. In particular, a proximal magnetic order can induce a gap in an otherwise linear spectrum of the Dirac electrons residing on the surface of a TI, resulting in a phase transition to a quantum anomalous Hall (QAH) state  \cite{Pankratov1987, *Liu2008, *Yu2010}. This QAH phase manifests itself via the appearance of quantized surface Hall currents and chiral modes (at the boundary where the gap vanishes), which, in turn, give rise to two types of magnetoelectric effects. Firstly, the Oersted fields originating from the surface Hall currents result in a topological magnetoelectric effect \cite{ZhangTME,*VanderTME} (a condensed matter realization of axion electrodynamics \cite{AED}). Examples of predicted consequences of this magnetoelectric effect include: appearance of image magnetic monopole \cite{Qi2009} and charging of magnetization textures \cite{Nomura2010}. Secondly, the surface Hall currents and chiral modes exert exchange coupling-induced spin torque on the magnetic order both in and out of equilibrium. Here, the magnetoelectric effects explored theoretically include: surface Hall currents-induced switching of monodomain magnets \cite{FranzPRL}; coupled dynamics of domain wall and chiral mode imprinted by the wall \cite{YaroslavPRL} \cite{Yago1, *Yago2}; and equilibrium spin torque-induced interfacial anisotropy and Dzyaloshinskii-Moriya interaction  \cite{SouchengPRL, *YaroslavPRBr}.

Technologically, the magnetoelectric effects in QAH regime provide an opportunity to construct spintronic devices with minimal dissipation, thanks to the dissipationless nature of Hall currents and chiral propagation. A prototypical example of a spintronic application envisioned for next generation information processing is based on a domain-wall, owing to its non-volatile nature. In such a device, information stored in a domain-wall configuration is manipulated electrically for memory (such as race-track memory \cite{Parkin11042008}) and logic \cite{Allwood2005} applications. However, mitigating Joule heating incurred during this electrical manipulation is one of the major challenges \cite{Parkin11042008}, demanding search for alternate mechanisms to control domain configurations electrically. A natural question then arises as to the possibility of taking advantage of low-dissipation magnetoelectric effects in the QAH phase for this purpose? In this Letter, we answer this question affirmatively by proposing a new energy-efficient scheme to move domain walls electrically, exploiting the magnetoelectric coupling mediated by the chiral modes in the QAH regime. 

\begin{figure}[h]
\centering
\includegraphics[width=0.5\textwidth]{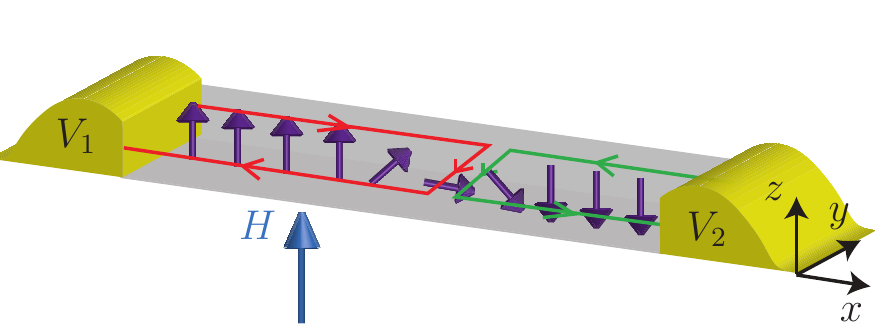}
\caption{Schematic of the magnetoelectric system: A magnetically doped topological insulator with the configuration of a domain wall connected by two electrical leads that can be maintained at voltages $V_1$ and $V_2$. The domain wall configuration results in opposite quantum anomalous Hall effects in two domains, with the presence of chiral edge modes as indicated by the red and green lines. An external magnetic field can be applied to move the domain wall.}
\label{dw_cartoon}
\end{figure}

\textit{Main results}.|The proposed system consists of a single domain wall, separating domains with spins oriented perpendicular (i.e. along $\pm z$) to the plane of the TI film (see Fig. ~\ref{dw_cartoon}). This system is motivated by recent experimental observation of QAH effect in magnetically doped TI \cite{Chang2013, *Xufeng2014} and giant spin-orbit torque-induced switching of perpendicular magnets \cite{Fan2014} 
in the same system. The opposite out-of-plane magnetizations, in the respective domains, open gaps of opposite signs. This results in ``local" QAH states with chiral edge modes propagating in the opposite directions on the same edge of the sample (shown by the red and green lines, respectively). At the domain wall, the out-of-plane component, and hence the gap, vanishes, imprinting two comoving chiral modes \cite{TI_rev_east}. The magnetoelectric system is then formed by connecting the film to two electrical leads as shown in Fig.~\ref{dw_cartoon}.

There are two main results of this Letter, summarized in Eqs.~(\ref{main1}) and (\ref{main2}). Firstly, the motion of the domain wall pumps charge current between the reservoirs maintained at the same electrochemical potential. For a domain wall moving with a velocity $v$ in a mirror (about the $xz$ plane) symmetric system, this pumped current is given by:
\ben
I_p = evk_FR/\pi.
\label{main1}
\een
Here, $e$ is the charge of an electron, $k_F$ is the Fermi wave number of electrons propagating along the chiral modes and $R$ is the reflection coefficient of the domain-wall region (from the view of the edge modes). This phenomenon is found to be similar to the particle current dragged by a piston, having a reflectivity $R$, moving through a gas of particles. Secondly, sending current through the chiral modes (for example by application of an external voltage, $V$, between the reservoirs) induces motion of the domain wall. The corresponding domain-wall velocity is given by $v= \gamma H_V \delta_w/\alpha$ with $\gamma$, $\delta_w$ and $\alpha$ being the gyromagnetic ratio, domain-wall width and Gilbert damping, respectively. Additionally, here, we have defined an electrical bias-induced effective magnetic field as:
\ben
H_V=\frac{\pi \theta_V \hbar I_q}{2e\lambda_F M_s d W}, 
\label{main2}
\een
with $I_q\equiv e^2V/h$ being the Landauer-B\"{u}ttiker current through a single mode and $\theta_V \equiv 4R$. Here, $\hbar$ is the reduced Planck's constant, $\lambda_F \equiv 2\pi/k_F$ is the Fermi wavelength, while $M_s$, $d$, and $W$ are the saturation magnetization, thickness, and the width of the magnetically-doped TI film, respectively. Recently, utilizing spin-orbit interaction has emerged as an efficient scheme for moving domain walls. In one popular scheme, the ferromagnet is interfaced with a heavy metal which, via a spin Hall-like mechanism, applies torque to induce domain-wall motion \cite{emorinm13, *ryunnt13}. The effective drive field for such a mechanism is given by: $H_S=\pi \theta_s \hbar I/2e t_m M_s d W$, where $t_m$ is the thickness of the heavy metal and the efficiency of the drive field is given by the so-called spin Hall angle $\theta_s$ (more precisely, referred to as the spin Hall tangent \cite{Sh_pheno}). Comparing $H_V$ with $H_S$ we note that $\theta_V$ plays the role of efficiency, much like the spin Hall angle. Moreover, we show that with increasing sample width, when the reflection coefficient approaches $0.5$, the effective spin Hall efficiency reaches the universal value of $\theta_V=2$, for the proposed mechanism. For typical heavy metals, $\theta_S \sim 0.1$ \cite{hoff_she}, suggesting the QAH regime could be an attractive alternative for efficient electrical manipulation of domains. Furthermore, dissipation in the QAH regime is localized only at the domain wall and the electrical contacts. Consequently, QAH phase has the advantage of dissipation scaling with the number of domain walls, as opposed to the entire  length of the magnetic sample outside of the QAH regime \cite{Parkin11042008}. 

\textit{Model}.|The magnetic sector, exchange-coupled to Dirac electrons on the surface of TI, is described by the free energy  \cite{YaroslavPRL}:
\ben
\mathcal{F}_M= \int (A |{\bf \nabla} {\bf m}|^2/2+Km_z^2/2+Dm_z{\bf \nabla} \cdot {\bf m}-{HM_sm_z}) d^3 r,
\label{free_en_m}
\een
where ${\bf m}\equiv(m_x,m_y,m_z)$ is a unit vector oriented along the magnetization, $H$ is a $z$-directed external magnetic field and we have defined ${\bf \nabla} \equiv {\bf x}\partial_x + {\bf y} \partial_y$. Phenomenologically, the parameters $A$ and $K$ represent the strength of exchange interaction and perpendicular anisotropy in the magnet, respectively, while $D$ parameterizes any inversion symmetry breaking-induced interfacial Dzyaloshinskii-Moriya interaction \cite{Dzioloshinskii, *Moriya} (for example due to different materials on top and bottom of the magnetic TI film). Microscopically, a finite $D$ is induced by equilibrium spin torques, arising from integrating out electrons exchange coupled to the magnetic texture \cite{YaroslavPRL}. 
We focus on the configuration of a one-dimensional domain wall and adopt the collective coordinate approach \cite{Thiele, *Bazaliy2008} to describe its dynamics. To this end, the free energy $\mathcal{F}_M$ can be expressed in terms of $X$ and $\Phi$ by substitution of a Walker wall ansatz \cite{walker1974}: $\ln\{\tan [\theta(x)/2]\}=(x-X)/\delta_w$; $\phi(x)=\Phi$ in Eq.~(\ref{free_en_m}). Here, we have parameterized the magnetization field in terms of the polar and azimuthal angles as: ${\bf m} \equiv (\sin \theta \cos \phi, \sin \theta \sin \phi, \cos \theta)$, while $X$ and $\Phi$ label, respectively, the position and the azimuthal angle of the domain-wall magnetization at the location where its out-of-plane component vanishes. Adding the electrical reservoirs, we then arrive at the free energy for our magnetoelectric system as:
\begin{eqnarray}
\mathcal{F}(Q,X,\Phi)= -2M_sdW(HX+H_D\delta_w \cos \Phi)-QV
.
\label{free_en}
\end{eqnarray}
Here, we have defined an effective Dzyaloshinskii-Moriya field $H_D\equiv \pi D/2M_s\delta_w$. The electronic degrees of freedom of the reservoirs are represented by a single thermodynamic variable $Q\equiv (Q_1-Q_2)/2$, with $Q_1$ and $Q_2$ denoting the electrical charge in reservoirs 1 and 2, respectively. The conjugate force, $V \equiv V_1-V_2$, is the voltage difference between the respective contacts. Such reduction to single variable $Q$ can be done assuming $\dot{Q_1}+\dot{Q_2}=0$ (i.e., no charge is being deposited in the QAH region), whose applicability will be discussed later.

Within the linear response, the out-of-equilibrium dynamics is then governed by \cite{landauBKOns}:
\ben
\begin{pmatrix}
\dot{Q}\\
\dot{X}\\
\dot{\Phi}
\end{pmatrix}=
\begin{pmatrix}
\mathcal{L}_{qq} & \mathcal{L}_{qx} & \mathcal{L}_{q\phi} \\
\mathcal{L}_{xq} & \mathcal{L}_{xx} & \mathcal{L}_{x\phi} \\
\mathcal{L}_{\Phi q} & \mathcal{L}_{\phi x} & \mathcal{L}_{\phi \phi} \\
\end{pmatrix}
\begin{pmatrix}
f_Q\\
f_X\\
f_\Phi
\end{pmatrix},
\label{lin_res}
\een
where the thermodynamically conjugate forces, identified from Eq.~(\ref{free_en})
, are $f_Q=V$, $f_X=2M_stWH$ and $f_\Phi=-2M_stWH_D\delta_w\sin\phi$. The kinetic coefficient for the electric sector, as given by the Ohm's law, is simply the conductance, $G$, i.e., $\mathcal{L}_{qq}=G$. For the magnetic sector, the kinetic coefficients are obtained from the substitution of Walker ansatz into the Landau-Lifshitz-Gilbert equation \cite{landauBK80}: $\partial_t {\bf m}=-\gamma {\bf m} \times \delta_{\bf m}\mathcal{F}_M+\alpha {\bf m} \times \partial_t{\bf m}$. Consequently, we have \cite{Thiaville12}: $\mathcal{L}_{xx}=(\gamma /2M_stW)\alpha \delta_w/(1+\alpha^2)$, $\mathcal{L}_{\phi \phi}=(\gamma /2M_stW)\alpha/\delta_w(1+\alpha^2)$ and the Onsager reciprocal pair $\mathcal{L}_{\phi x}=-\mathcal{L}_{x\phi}=(\gamma /2M_stW)/(1+\alpha^2)$. Finally, the magnetoelectric coupling is encoded in $\mathcal{L}_{qx}$, $\mathcal{L}_{q\phi}$| describing domain-wall dynamics-induced charge pumping; and $\mathcal{L}_{xq}$, $\mathcal{L}_{\phi q}$| describing the electrical bias-induced domain-wall motion.

\begin{figure}[h]
\centering
\includegraphics[width=0.5\textwidth]{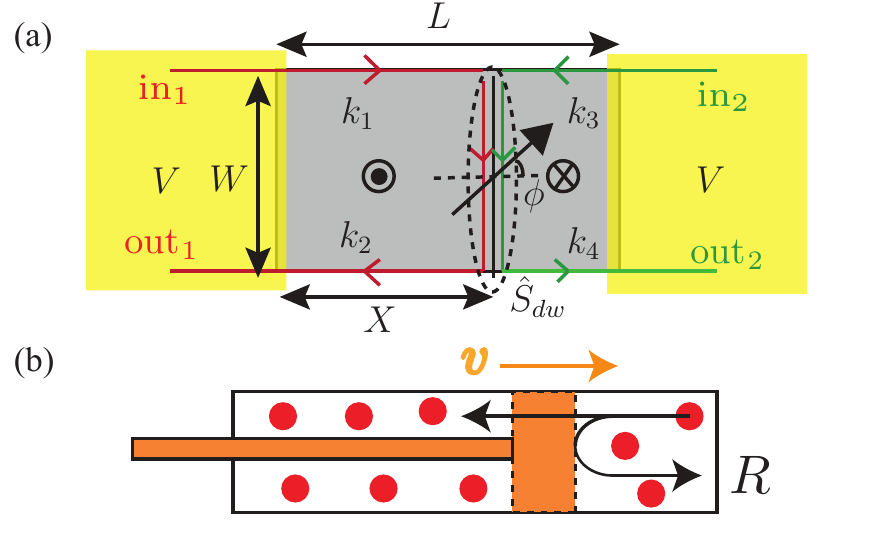}
\caption{Schematic depicting the scattering process. (a) Top view of the QAH system of width $W$ and length $L$, with a domain wall at a distance $X$ from the left contact. The magnetization at the domain wall center is depicted by the arrow making an angle $\phi$ with the $x$ axis. Electrical transport occurs via chiral modes at the edges of the sample (with their respective wave vectors denoted by $k_i$), which upon reaching the domain-wall region (marked by the dashed ellipse) can intermix. This intermixing is parameterized by a scattering matrix of the wall region $\hat{S}_{dw}$. (b) Schematic of ``1D piston" with the gas of particles (represented by the filled circles) constrained to move along the length of the wire. The piston, having an opacity $R$, is moving with a velocity $v$ through the gas of the particles and can thus drag a flux. This phenomenon, as discussed in the text, is equivalent to the current pumped by the moving domain wall in (a).} 
\label{scattering}
\end{figure}

\textit{Charge transport and pumping}.|The charge pumping by a dynamic domain wall is an example of a parametric pump. The electrical transport between the reservoirs occur via the chiral modes as sketched in Fig.~\ref{scattering}(a). The modes emanating from the respective contacts reach the domain-wall region without scattering where, via a finite probability to intermix, they can either transmit or reflect. The scattering, in turn, depends parametrically on the location of the domain wall via the propagation phases acquired during the journey of chiral modes between the reservoirs and the domain wall. Thus a domain wall moving with a constant velocity, according to the B\"{u}ttiker-Brouwer's formalism \cite{Buttikerpump, *brouwer98}, results in the pumping of a charge current. To this end, we start by writing down the scattering matrix, $\hat{S}$, for the QAH insulator with a domain wall located at a distance $X$ from the left contact. Adopting the following definition for scattering matrix $\bigl(\begin{smallmatrix}
\rm{out}_1\\ \rm{out}_2 
\end{smallmatrix} \bigr)
\equiv
\hat{S}
\bigl(\begin{smallmatrix}
\rm{in}_1\\ \rm{in}_2 
\end{smallmatrix} \bigr)$, where $\rm{in}_1(\rm{out}_1)$ and $\rm{in}_2(\rm{out}_2)$ represent the amplitude of incoming (outgoing) modes from left and right contacts, respectively,  we have:
\ben
\hat{S}=
\begin{pmatrix}
  re^{i(k_1+k_2)X} & t^\prime e^{ik_3L}e^{i(k_2-k_3)X} \\
  {t}e^{ik_4L}e^{i(k_1-k_4)X} & r^\prime e^{i(k_3+k_4)L}e^{-i(k_3+k_4)X}
\end{pmatrix}.
\label{sm}
\een 
Here, $L$ is the length of the QAH insulator between the reservoirs, $k_i$ represent the propagation wave vectors in different sections of the chiral modes and $\hat{S}_{dw} \equiv \bigl(\begin{smallmatrix}
r&t^\prime\\ t &r^\prime
\end{smallmatrix} \bigr)$ denotes the scattering matrix of the wall region, which can depend on $\phi$, as depicted in Fig.~\ref{scattering}. The parametric dependence of scattering matrix (\ref{sm}) on the domain-wall position and angle then dictates that a dynamic domain wall pumps current \cite{Buttikerpump, *brouwer98}:
\ben
I_m(T)=\frac{e}{2\pi}\Im \sum_\beta \Big[\Big(\frac{\partial S_{m\beta}}{\partial X}S^*_{m\beta}\Big)\dot{X}+ \Big(\frac{\partial S_{m\beta}}{\partial \phi}S^*_{m\beta}\Big)\dot{\Phi}\Big]. 
\label{BB}
\een
Here, $I_m(T)$ is the net charge current emitted from the contact $m$ at time $T$, $e$ is the charge of an electron, $S_{ij}$ is the matrix element of the scattering matrix $\hat{S}$ and $\Im$ stands for the imaginary part. 

In this Letter, we are interested in the charge pumped by a steady state of the domain wall moving with a constant velocity $v$. This regime is established when the strength of the external magnetic field is below the so-called Walker breakdown field, i.e., $H<H_W$ \cite{walker1974}. Here, from the equation of motion for the magnetic angle in Eq.~(\ref{lin_res}), this Walker breakdown field is, in turn, set by the Dzyaloshinskii-Moriya field as $H_W=\alpha H_D$. In this regime $\dot{\Phi}=0$ and $v=\gamma H\delta_w/\alpha$, resulting in the  pumped charge current, obtained upon substitution of the scattering matrix from (\ref{sm}) in Eq.~\ref{BB}, as:
\begin{align}
I_1 &=\frac{\gamma e H \delta_w}{2\pi \alpha}(R\{k_1+k_3\}+k_2-k_3), \nonumber \\
I_2 &=-\frac{\gamma e H \delta_w}{2\pi \alpha}(R\{k_1+k_3\}+k_4-k_1),
\label{cp_general}
\end{align}
where $R\equiv |r|^2$ is the reflection probability of the domain-wall region. We note that, in general, $I_1+I_2 \neq 0$, reflecting the fact that charge can be repopulated along the edges, adding a capacitive energy cost to the free energy of our magnetoelectric system \cite{Buttikerpump, *brouwer98}. However, focusing, for clarity, on systems which are mirror (about the $xz$ plane) symmetric results in $I_1+I_2 = 0$, as explained next. 

First, we note that a simultaneous mirror (about the $xz$ plane) and time reversal transformation leaves the magnetic configuration and external fields invariant while interchanging $k_1$ with $k_2$ (and $k_3$ with $k_4$). Consequently for the aforementioned mirror-symmetric case, which is satisfied in technologically relevant nanowire geometry of racetracks \cite{Parkin11042008}, we have $k_1=k_2$ and $k_3=k_4$. Additionally, we note that the two domains and corresponding chiral modes are related by time reversal, invoking which, we further get $k_1=k_3$ and $k_2=k_4$ \footnote{Although, in principle, the fixed external magnetic field breaks the time-reversal symmetry, we are interested in the regime where the associated Zeeman energy scale is much smaller than the exchange-correlation energy. In this case, the dispersion relations for chiral modes are not significantly affected by the external field}. Under these assumptions, substituting $k_1=k_2=k_3=k_4=k_F$ in Eq.~(\ref{cp_general}), we arrive at the first main result of the letter: $I_1=-I_2=I_p$, where the pumped current $I_p$ is given by Eq.~(\ref{main1}). This pumped current can also be understood in terms of charge pushed by a piston, having a reflection coefficient $R$, moving with a velocity $v$ in a strictly one-dimensional wire [referred here as the ``1D piston" and sketched in Fig.~\ref{scattering}(b)]. Taking, for simplicity, a Galilean invariant 1-D piston with quadratic energy dispersion, i.e. $E(k)=\hbar^2 k^2/2m$ (with $m$ being the effective mass of fermionic particles at low temperature), the particle current dragged by the piston can be calculated by going into the frame of reference where the piston is static. In this frame, the wave vectors get boosted as : $k \rightarrow k+\delta k$ with $\delta k=mv/\hbar$. Consequently, the energy difference between the left and the right movers, i.e., $\Delta E=2(\partial E/\partial k) \delta k$, appears as an effective bias voltage given by $eV_{\rm eff}=\Delta E=2\hbar k_F v$. Given the reflectivity of the piston is $R$, the Landauer-B\"{u}ttiker charge current in the frame of reference of the piston is then given by $I_p=(R-1)e^2V_{\rm eff}/h=(R-1)evk_F/\pi$. Finally, transforming this current back in the wire frame, the pumped current becomes $I=Revk_F/\pi$, establishing the equivalence mentioned above. Having found the charge pumped by a moving domain wall, we get $\mathcal{L}_{qx}=\gamma e \delta_wR/\alpha M_s dW$ using Eq.~(\ref{main1}) and Eq.~(\ref{lin_res}). Invoking Onsager reciprocity \cite{landauBKOns} we then arrive at the other main result with the electrical bias-induced effective field given by Eq.~(\ref{main2}). 

\begin{figure}[h]
\centering
\includegraphics[scale=0.70]{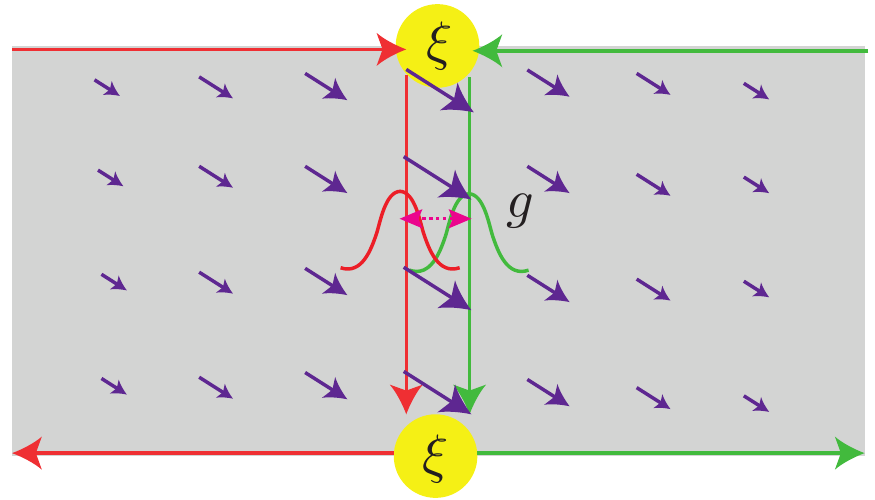}
\caption{Depiction of the intermixing process in the domain-wall region: The red and the green channel, i.e. the modes emanated from the left and right contacts of Fig.~\ref{dw_cartoon}, can intermix upon reaching the domain-wall region. This intermixing is facilitated by two processes: (i) sharp turns in the conduction path at either ends of the domain wall, represented by circular scatterers with transmission probability $\xi$. (ii) during propagation along the domain wall tunneling can occur between the modes due to finite width of each mode (as depicted by the overlap of two Gaussians), which is parameterized by the tunneling conductance per unit length $g$.}
\label{micro}
\end{figure} 

Microscopically, the reflection coefficient governing the efficiency of this effective field is determined by intermixing of the chiral modes at the domain-wall region. Recently, such intermixing process has been discussed in the context of edge state mixing in the quantum Hall state of graphene \cite{So_graphene}. There are two sources of this intermixing process which are illustrated schematically in Fig.~\ref{micro}. (a) The potentially abrupt change in the direction of propagation at the beginning and the end of the domain wall (as indicated by the circular scatterer) can result in distribution of charges between the chiral modes. Due to conservation of current, this distribution can be parameterized by a single parameter $\xi$, denoting the transmission probability of each such scatterer. (b) During the propagation of modes along the domain wall, proximity-induced finite tunneling (as parameterized by a tunneling conductance per unit length $g$), can also result in the equilibration process. The reflection coefficient is  then given by \cite{So_graphene}:
\ben
R=\frac{1+(1-2\xi)^2e^{-W/\lambda}}{2},
\een
where $\lambda \equiv e^2/2gh$ is defined as the equilibration length for intermode scattering during propagation along the domain wall. When the full equilibration between the chiral modes occurs by scattering at the start and end of the domain wall, i.e., for $\xi=1/2$, the reflection coefficient becomes independent of the sample width, just as the spin Hall angle, and takes a value of $R=1/2$. On the other hand, in the regime when the intermixing at the sharp turns can be neglected, i.e. $\xi=0$, the width of the sample as compared to equilibration length becomes an additional important parameter governing $R$. For $W \gg \lambda$, the reflection coefficient again approaches the value of $R=1/2$. Consequently, $\theta_V$ for the proposed mechanism can quite generally be expected to approach a universal value of 2.

\textit{Quantitative estimates}|
For quantitative estimates we focus on magnetic-TI structures \cite{Chang2013, *Xufeng2014} with material parameters based on Cr-doped (Bi,Sb)$_2$Te$_3$ system. The exchange parameter is estimated as $A \sim k_BT_c/a \sim 4 \times 10^{-8}$~ erg/cm, where $T_c=30$~K is the Curie temperature and $a \sim 1$~nm is the typical distance between two Cr atoms, while $K=5 \times 10^4$~erg/cm$^3$ and $M_s=10$~emu/cm$^3$. Focusing first on charge pumping, these material parameters result in a domain-wall velocity of $v=\gamma H \delta_w/\alpha \sim 100$~ m/sec for typical external fields of strength $H= 10$ Oe and damping parameter of $\alpha \sim 0.01$. Here we have used the domain-wall widths $\delta_w \sim \sqrt{A/K} \sim 10$~nm. For estimating the largest possible pumped current ($I_{\rm max}$), while still maintaining our low energy description based on the QAH insulator phase, we have $k_F \sim \Delta/\hbar v_F$. Here, $v_F$ parameterizes the linear dispersion of the form $E(k)=\hbar v_F k$ for the chiral modes, and $2\Delta$ is the gap opened in the surface states dispersion due to exchange coupling to magnetic order. Using this, we get $I_{\rm max} \sim eRv\Delta/hv_F$. The Fermi velocity for the chiral modes is the most uncertain parameter here, estimating which from the typical surface state Fermi velocity $v_F=10^5$~m/sec \cite{Zhang2009}, and using  $\Delta \sim 0.1$~eV from ARPES measurements \cite{Xufeng2012}, we get $I_{\rm max} \sim 4$~nA, which is within the reach of experimental observation \footnote{The underlying adiabatic approximation, used in deriving the pumping, is valid for values of $v$ such that the effective bias in domain wall's reference frame $V_{\rm eff}$ (as discussed in the main text) is much less than the gap $2\Delta$. Substituting for $V_{\rm eff}$ from the main text this becomes equivalent to $v \ll v_F$, which is satisfied for the chosen material parameters.}. Next, for current-induced domain-wall motion, the effective magnetic field for a bias of $eV=\Delta=0.1$~eV and a typical nanowire geometry with $W=100$~nm and $t=5$~nm is estimated to be $H_V=2\pi R\hbar I_q/e\lambda_F M_stW \sim 1$ Oe. Consequently, expected order of current-induced domain-wall velocity is given by $v \sim 10$~m/sec. This value is comparable to the spin-orbit torque-induced velocity in heavy metals for about an order larger current density of $J \sim 10^7$~ A/cm$^2$ \cite{emorinm13}, reflecting the higher efficiency of the proposed mechanism. 

\textit{Acknowledgments}| This work was supported by FAME (an SRC STARnet center sponsored by MARCO and DARPA). Discussions with So Takei, Yabin Fan, and Kang Wang are gratefully acknowledged.


\begin{thebibliography}{38}%
\makeatletter
\providecommand \@ifxundefined [1]{%
 \@ifx{#1\undefined}
}%
\providecommand \@ifnum [1]{%
 \ifnum #1\expandafter \@firstoftwo
 \else \expandafter \@secondoftwo
 \fi
}%
\providecommand \@ifx [1]{%
 \ifx #1\expandafter \@firstoftwo
 \else \expandafter \@secondoftwo
 \fi
}%
\providecommand \natexlab [1]{#1}%
\providecommand \enquote  [1]{``#1''}%
\providecommand \bibnamefont  [1]{#1}%
\providecommand \bibfnamefont [1]{#1}%
\providecommand \citenamefont [1]{#1}%
\providecommand \href@noop [0]{\@secondoftwo}%
\providecommand \href [0]{\begingroup \@sanitize@url \@href}%
\providecommand \@href[1]{\@@startlink{#1}\@@href}%
\providecommand \@@href[1]{\endgroup#1\@@endlink}%
\providecommand \@sanitize@url [0]{\catcode `\\12\catcode `\$12\catcode
  `\&12\catcode `\#12\catcode `\^12\catcode `\_12\catcode `\%12\relax}%
\providecommand \@@startlink[1]{}%
\providecommand \@@endlink[0]{}%
\providecommand \url  [0]{\begingroup\@sanitize@url \@url }%
\providecommand \@url [1]{\endgroup\@href {#1}{\urlprefix }}%
\providecommand \urlprefix  [0]{URL }%
\providecommand \Eprint [0]{\href }%
\providecommand \doibase [0]{http://dx.doi.org/}%
\providecommand \selectlanguage [0]{\@gobble}%
\providecommand \bibinfo  [0]{\@secondoftwo}%
\providecommand \bibfield  [0]{\@secondoftwo}%
\providecommand \translation [1]{[#1]}%
\providecommand \BibitemOpen [0]{}%
\providecommand \bibitemStop [0]{}%
\providecommand \bibitemNoStop [0]{.\EOS\space}%
\providecommand \EOS [0]{\spacefactor3000\relax}%
\providecommand \BibitemShut  [1]{\csname bibitem#1\endcsname}%
\let\auto@bib@innerbib\@empty
\bibitem [{\citenamefont {Moore}(2010)}]{Moore2010}%
  \BibitemOpen
  \bibfield  {author} {\bibinfo {author} {\bibfnamefont {J.~E.}\ \bibnamefont
  {Moore}},\ }\href {\doibase 10.1038/nature08916} {\bibfield  {journal}
  {\bibinfo  {journal} {Nature}\ }\textbf {\bibinfo {volume} {464}},\ \bibinfo
  {pages} {194} (\bibinfo {year} {2010})}\BibitemShut {NoStop}%
\bibitem [{\citenamefont {Hasan}\ and\ \citenamefont
  {Kane}(2010)}]{TI_rev_east}%
  \BibitemOpen
  \bibfield  {author} {\bibinfo {author} {\bibfnamefont {M.~Z.}\ \bibnamefont
  {Hasan}}\ and\ \bibinfo {author} {\bibfnamefont {C.~L.}\ \bibnamefont
  {Kane}},\ }\href {\doibase 10.1103/RevModPhys.82.3045} {\bibfield  {journal}
  {\bibinfo  {journal} {Rev. Mod. Phys.}\ }\textbf {\bibinfo {volume} {82}},\
  \bibinfo {pages} {3045} (\bibinfo {year} {2010})}\BibitemShut {NoStop}%
\bibitem [{\citenamefont {Qi}\ and\ \citenamefont {Zhang}(2011)}]{TI_rev_west}%
  \BibitemOpen
  \bibfield  {author} {\bibinfo {author} {\bibfnamefont {X.-L.}\ \bibnamefont
  {Qi}}\ and\ \bibinfo {author} {\bibfnamefont {S.-C.}\ \bibnamefont {Zhang}},\
  }\href {\doibase 10.1103/RevModPhys.83.1057} {\bibfield  {journal} {\bibinfo
  {journal} {Rev. Mod. Phys.}\ }\textbf {\bibinfo {volume} {83}},\ \bibinfo
  {pages} {1057} (\bibinfo {year} {2011})}\BibitemShut {NoStop}%
\bibitem [{\citenamefont {Pankratov}(1987)}]{Pankratov1987}%
  \BibitemOpen
  \bibfield  {author} {\bibinfo {author} {\bibfnamefont {O.}~\bibnamefont
  {Pankratov}},\ }\href {\doibase 10.1016/0375-9601(87)90307-0} {\bibfield
  {journal} {\bibinfo  {journal} {Physics Letters A}\ }\textbf {\bibinfo
  {volume} {121}},\ \bibinfo {pages} {360} (\bibinfo {year}
  {1987})}\BibitemShut {NoStop}%
\bibitem [{\citenamefont {Liu}\ \emph {et~al.}(2008)\citenamefont {Liu},
  \citenamefont {Qi}, \citenamefont {Dai}, \citenamefont {Fang},\ and\
  \citenamefont {Zhang}}]{Liu2008}%
  \BibitemOpen
  \bibfield  {author} {\bibinfo {author} {\bibfnamefont {C.-X.}\ \bibnamefont
  {Liu}}, \bibinfo {author} {\bibfnamefont {X.-L.}\ \bibnamefont {Qi}},
  \bibinfo {author} {\bibfnamefont {X.}~\bibnamefont {Dai}}, \bibinfo {author}
  {\bibfnamefont {Z.}~\bibnamefont {Fang}}, \ and\ \bibinfo {author}
  {\bibfnamefont {S.-C.}\ \bibnamefont {Zhang}},\ }\href {\doibase
  10.1103/PhysRevLett.101.146802} {\bibfield  {journal} {\bibinfo  {journal}
  {Phys. Rev. Lett.}\ }\textbf {\bibinfo {volume} {101}},\ \bibinfo {pages}
  {146802} (\bibinfo {year} {2008})}\BibitemShut {NoStop}%
\bibitem [{\citenamefont {Yu}\ \emph {et~al.}(2010)\citenamefont {Yu},
  \citenamefont {Onose}, \citenamefont {Kanazawa}, \citenamefont {Park},
  \citenamefont {Han}, \citenamefont {Matsui}, \citenamefont {Nagaosa},\ and\
  \citenamefont {Tokura}}]{Yu2010}%
  \BibitemOpen
  \bibfield  {author} {\bibinfo {author} {\bibfnamefont {X.~Z.}\ \bibnamefont
  {Yu}}, \bibinfo {author} {\bibfnamefont {Y.}~\bibnamefont {Onose}}, \bibinfo
  {author} {\bibfnamefont {N.}~\bibnamefont {Kanazawa}}, \bibinfo {author}
  {\bibfnamefont {J.~H.}\ \bibnamefont {Park}}, \bibinfo {author}
  {\bibfnamefont {J.~H.}\ \bibnamefont {Han}}, \bibinfo {author} {\bibfnamefont
  {Y.}~\bibnamefont {Matsui}}, \bibinfo {author} {\bibfnamefont
  {N.}~\bibnamefont {Nagaosa}}, \ and\ \bibinfo {author} {\bibfnamefont
  {Y.}~\bibnamefont {Tokura}},\ }\href {\doibase 10.1038/nature09124}
  {\bibfield  {journal} {\bibinfo  {journal} {Nature}\ }\textbf {\bibinfo
  {volume} {465}},\ \bibinfo {pages} {901} (\bibinfo {year}
  {2010})}\BibitemShut {NoStop}%
\bibitem [{\citenamefont {Qi}\ \emph {et~al.}(2008)\citenamefont {Qi},
  \citenamefont {Hughes},\ and\ \citenamefont {Zhang}}]{ZhangTME}%
  \BibitemOpen
  \bibfield  {author} {\bibinfo {author} {\bibfnamefont {X.-L.}\ \bibnamefont
  {Qi}}, \bibinfo {author} {\bibfnamefont {T.~L.}\ \bibnamefont {Hughes}}, \
  and\ \bibinfo {author} {\bibfnamefont {S.-C.}\ \bibnamefont {Zhang}},\ }\href
  {\doibase 10.1103/PhysRevB.78.195424} {\bibfield  {journal} {\bibinfo
  {journal} {Phys. Rev. B}\ }\textbf {\bibinfo {volume} {78}},\ \bibinfo
  {pages} {195424} (\bibinfo {year} {2008})}\BibitemShut {NoStop}%
\bibitem [{\citenamefont {Essin}\ \emph {et~al.}(2009)\citenamefont {Essin},
  \citenamefont {Moore},\ and\ \citenamefont {Vanderbilt}}]{VanderTME}%
  \BibitemOpen
  \bibfield  {author} {\bibinfo {author} {\bibfnamefont {A.~M.}\ \bibnamefont
  {Essin}}, \bibinfo {author} {\bibfnamefont {J.~E.}\ \bibnamefont {Moore}}, \
  and\ \bibinfo {author} {\bibfnamefont {D.}~\bibnamefont {Vanderbilt}},\
  }\href {\doibase 10.1103/PhysRevLett.102.146805} {\bibfield  {journal}
  {\bibinfo  {journal} {Phys. Rev. Lett.}\ }\textbf {\bibinfo {volume} {102}},\
  \bibinfo {pages} {146805} (\bibinfo {year} {2009})}\BibitemShut {NoStop}%
\bibitem [{\citenamefont {Wilczek}(1987)}]{AED}%
  \BibitemOpen
  \bibfield  {author} {\bibinfo {author} {\bibfnamefont {F.}~\bibnamefont
  {Wilczek}},\ }\href {\doibase 10.1103/PhysRevLett.58.1799} {\bibfield
  {journal} {\bibinfo  {journal} {Phys. Rev. Lett.}\ }\textbf {\bibinfo
  {volume} {58}},\ \bibinfo {pages} {1799} (\bibinfo {year}
  {1987})}\BibitemShut {NoStop}%
\bibitem [{\citenamefont {Qi}\ \emph {et~al.}(2009)\citenamefont {Qi},
  \citenamefont {Li}, \citenamefont {Zang},\ and\ \citenamefont
  {Zhang}}]{Qi2009}%
  \BibitemOpen
  \bibfield  {author} {\bibinfo {author} {\bibfnamefont {X.-L.}\ \bibnamefont
  {Qi}}, \bibinfo {author} {\bibfnamefont {R.}~\bibnamefont {Li}}, \bibinfo
  {author} {\bibfnamefont {J.}~\bibnamefont {Zang}}, \ and\ \bibinfo {author}
  {\bibfnamefont {S.-C.}\ \bibnamefont {Zhang}},\ }\href {\doibase
  10.1126/science.1167747} {\bibfield  {journal} {\bibinfo  {journal}
  {Science}\ }\textbf {\bibinfo {volume} {323}},\ \bibinfo {pages} {1184}
  (\bibinfo {year} {2009})}\BibitemShut {NoStop}%
\bibitem [{\citenamefont {Nomura}\ and\ \citenamefont
  {Nagaosa}(2010)}]{Nomura2010}%
  \BibitemOpen
  \bibfield  {author} {\bibinfo {author} {\bibfnamefont {K.}~\bibnamefont
  {Nomura}}\ and\ \bibinfo {author} {\bibfnamefont {N.}~\bibnamefont
  {Nagaosa}},\ }\href {\doibase 10.1103/PhysRevB.82.161401} {\bibfield
  {journal} {\bibinfo  {journal} {Phys. Rev. B}\ }\textbf {\bibinfo {volume}
  {82}},\ \bibinfo {pages} {161401} (\bibinfo {year} {2010})}\BibitemShut
  {NoStop}%
\bibitem [{\citenamefont {Garate}\ and\ \citenamefont
  {Franz}(2010)}]{FranzPRL}%
  \BibitemOpen
  \bibfield  {author} {\bibinfo {author} {\bibfnamefont {I.}~\bibnamefont
  {Garate}}\ and\ \bibinfo {author} {\bibfnamefont {M.}~\bibnamefont {Franz}},\
  }\href {\doibase 10.1103/PhysRevLett.104.146802} {\bibfield  {journal}
  {\bibinfo  {journal} {Phys. Rev. Lett.}\ }\textbf {\bibinfo {volume} {104}},\
  \bibinfo {pages} {146802} (\bibinfo {year} {2010})}\BibitemShut {NoStop}%
\bibitem [{\citenamefont {Tserkovnyak}\ and\ \citenamefont
  {Loss}(2012)}]{YaroslavPRL}%
  \BibitemOpen
  \bibfield  {author} {\bibinfo {author} {\bibfnamefont {Y.}~\bibnamefont
  {Tserkovnyak}}\ and\ \bibinfo {author} {\bibfnamefont {D.}~\bibnamefont
  {Loss}},\ }\href {\doibase 10.1103/PhysRevLett.108.187201} {\bibfield
  {journal} {\bibinfo  {journal} {Phys. Rev. Lett.}\ }\textbf {\bibinfo
  {volume} {108}},\ \bibinfo {pages} {187201} (\bibinfo {year}
  {2012})}\BibitemShut {NoStop}%
\bibitem [{\citenamefont {Ferreiros}\ \emph {et~al.}(2015)\citenamefont
  {Ferreiros}, \citenamefont {Buijnsters},\ and\ \citenamefont
  {Katsnelson}}]{Yago1}%
  \BibitemOpen
  \bibfield  {author} {\bibinfo {author} {\bibfnamefont {Y.}~\bibnamefont
  {Ferreiros}}, \bibinfo {author} {\bibfnamefont {F.~J.}\ \bibnamefont
  {Buijnsters}}, \ and\ \bibinfo {author} {\bibfnamefont {M.~I.}\ \bibnamefont
  {Katsnelson}},\ }\href {\doibase 10.1103/PhysRevB.92.085416} {\bibfield
  {journal} {\bibinfo  {journal} {Phys. Rev. B}\ }\textbf {\bibinfo {volume}
  {92}},\ \bibinfo {pages} {085416} (\bibinfo {year} {2015})}\BibitemShut
  {NoStop}%
\bibitem [{\citenamefont {Ferreiros}\ and\ \citenamefont
  {Cortijo}(2014)}]{Yago2}%
  \BibitemOpen
  \bibfield  {author} {\bibinfo {author} {\bibfnamefont {Y.}~\bibnamefont
  {Ferreiros}}\ and\ \bibinfo {author} {\bibfnamefont {A.}~\bibnamefont
  {Cortijo}},\ }\href {\doibase 10.1103/PhysRevB.89.024413} {\bibfield
  {journal} {\bibinfo  {journal} {Phys. Rev. B}\ }\textbf {\bibinfo {volume}
  {89}},\ \bibinfo {pages} {024413} (\bibinfo {year} {2014})}\BibitemShut
  {NoStop}%
\bibitem [{\citenamefont {Zhu}\ \emph {et~al.}(2011)\citenamefont {Zhu},
  \citenamefont {Yao}, \citenamefont {Zhang},\ and\ \citenamefont
  {Chang}}]{SouchengPRL}%
  \BibitemOpen
  \bibfield  {author} {\bibinfo {author} {\bibfnamefont {J.-J.}\ \bibnamefont
  {Zhu}}, \bibinfo {author} {\bibfnamefont {D.-X.}\ \bibnamefont {Yao}},
  \bibinfo {author} {\bibfnamefont {S.-C.}\ \bibnamefont {Zhang}}, \ and\
  \bibinfo {author} {\bibfnamefont {K.}~\bibnamefont {Chang}},\ }\href
  {\doibase 10.1103/PhysRevLett.106.097201} {\bibfield  {journal} {\bibinfo
  {journal} {Phys. Rev. Lett.}\ }\textbf {\bibinfo {volume} {106}},\ \bibinfo
  {pages} {097201} (\bibinfo {year} {2011})}\BibitemShut {NoStop}%
\bibitem [{\citenamefont {Tserkovnyak}\ \emph {et~al.}(2015)\citenamefont
  {Tserkovnyak}, \citenamefont {Pesin},\ and\ \citenamefont
  {Loss}}]{YaroslavPRBr}%
  \BibitemOpen
  \bibfield  {author} {\bibinfo {author} {\bibfnamefont {Y.}~\bibnamefont
  {Tserkovnyak}}, \bibinfo {author} {\bibfnamefont {D.~A.}\ \bibnamefont
  {Pesin}}, \ and\ \bibinfo {author} {\bibfnamefont {D.}~\bibnamefont {Loss}},\
  }\href {\doibase 10.1103/PhysRevB.91.041121} {\bibfield  {journal} {\bibinfo
  {journal} {Phys. Rev. B}\ }\textbf {\bibinfo {volume} {91}},\ \bibinfo
  {pages} {041121} (\bibinfo {year} {2015})}\BibitemShut {NoStop}%
\bibitem [{\citenamefont {Parkin}\ \emph {et~al.}(2008)\citenamefont {Parkin},
  \citenamefont {Hayashi},\ and\ \citenamefont {Thomas}}]{Parkin11042008}%
  \BibitemOpen
  \bibfield  {author} {\bibinfo {author} {\bibfnamefont {S.~S.~P.}\
  \bibnamefont {Parkin}}, \bibinfo {author} {\bibfnamefont {M.}~\bibnamefont
  {Hayashi}}, \ and\ \bibinfo {author} {\bibfnamefont {L.}~\bibnamefont
  {Thomas}},\ }\href {\doibase 10.1126/science.1145799} {\bibfield  {journal}
  {\bibinfo  {journal} {Science}\ }\textbf {\bibinfo {volume} {320}},\ \bibinfo
  {pages} {190} (\bibinfo {year} {2008})}\BibitemShut {NoStop}%
\bibitem [{\citenamefont {Allwood}(2005)}]{Allwood2005}%
  \BibitemOpen
  \bibfield  {author} {\bibinfo {author} {\bibfnamefont {D.~A.}\ \bibnamefont
  {Allwood}},\ }\href {\doibase 10.1126/science.1108813} {\bibfield  {journal}
  {\bibinfo  {journal} {Science}\ }\textbf {\bibinfo {volume} {309}},\ \bibinfo
  {pages} {1688} (\bibinfo {year} {2005})}\BibitemShut {NoStop}%
\bibitem [{\citenamefont {Chang}\ \emph {et~al.}(2013)\citenamefont {Chang},
  \citenamefont {Zhang}, \citenamefont {Feng}, \citenamefont {Shen},
  \citenamefont {Zhang}, \citenamefont {Guo}, \citenamefont {Li}, \citenamefont
  {Ou}, \citenamefont {Wei}, \citenamefont {Wang}, \citenamefont {Ji},
  \citenamefont {Feng}, \citenamefont {Ji}, \citenamefont {Chen}, \citenamefont
  {Jia}, \citenamefont {Dai}, \citenamefont {Fang}, \citenamefont {Zhang},
  \citenamefont {He}, \citenamefont {Wang}, \citenamefont {Lu}, \citenamefont
  {Ma},\ and\ \citenamefont {Xue}}]{Chang2013}%
  \BibitemOpen
  \bibfield  {author} {\bibinfo {author} {\bibfnamefont {C.-Z.}\ \bibnamefont
  {Chang}}, \bibinfo {author} {\bibfnamefont {J.}~\bibnamefont {Zhang}},
  \bibinfo {author} {\bibfnamefont {X.}~\bibnamefont {Feng}}, \bibinfo {author}
  {\bibfnamefont {J.}~\bibnamefont {Shen}}, \bibinfo {author} {\bibfnamefont
  {Z.}~\bibnamefont {Zhang}}, \bibinfo {author} {\bibfnamefont
  {M.}~\bibnamefont {Guo}}, \bibinfo {author} {\bibfnamefont {K.}~\bibnamefont
  {Li}}, \bibinfo {author} {\bibfnamefont {Y.}~\bibnamefont {Ou}}, \bibinfo
  {author} {\bibfnamefont {P.}~\bibnamefont {Wei}}, \bibinfo {author}
  {\bibfnamefont {L.-L.}\ \bibnamefont {Wang}}, \bibinfo {author}
  {\bibfnamefont {Z.-Q.}\ \bibnamefont {Ji}}, \bibinfo {author} {\bibfnamefont
  {Y.}~\bibnamefont {Feng}}, \bibinfo {author} {\bibfnamefont {S.}~\bibnamefont
  {Ji}}, \bibinfo {author} {\bibfnamefont {X.}~\bibnamefont {Chen}}, \bibinfo
  {author} {\bibfnamefont {J.}~\bibnamefont {Jia}}, \bibinfo {author}
  {\bibfnamefont {X.}~\bibnamefont {Dai}}, \bibinfo {author} {\bibfnamefont
  {Z.}~\bibnamefont {Fang}}, \bibinfo {author} {\bibfnamefont {S.-C.}\
  \bibnamefont {Zhang}}, \bibinfo {author} {\bibfnamefont {K.}~\bibnamefont
  {He}}, \bibinfo {author} {\bibfnamefont {Y.}~\bibnamefont {Wang}}, \bibinfo
  {author} {\bibfnamefont {L.}~\bibnamefont {Lu}}, \bibinfo {author}
  {\bibfnamefont {X.-C.}\ \bibnamefont {Ma}}, \ and\ \bibinfo {author}
  {\bibfnamefont {Q.-K.}\ \bibnamefont {Xue}},\ }\href {\doibase
  10.1126/science.1234414} {\bibfield  {journal} {\bibinfo  {journal}
  {Science}\ }\textbf {\bibinfo {volume} {340}},\ \bibinfo {pages} {167}
  (\bibinfo {year} {2013})}\BibitemShut {NoStop}%
\bibitem [{\citenamefont {Kou}\ \emph {et~al.}(2014)\citenamefont {Kou},
  \citenamefont {Guo}, \citenamefont {Fan}, \citenamefont {Pan}, \citenamefont
  {Lang}, \citenamefont {Jiang}, \citenamefont {Shao}, \citenamefont {Nie},
  \citenamefont {Murata}, \citenamefont {Tang}, \citenamefont {Wang},
  \citenamefont {He}, \citenamefont {Lee}, \citenamefont {Lee},\ and\
  \citenamefont {Wang}}]{Xufeng2014}%
  \BibitemOpen
  \bibfield  {author} {\bibinfo {author} {\bibfnamefont {X.}~\bibnamefont
  {Kou}}, \bibinfo {author} {\bibfnamefont {S.-T.}\ \bibnamefont {Guo}},
  \bibinfo {author} {\bibfnamefont {Y.}~\bibnamefont {Fan}}, \bibinfo {author}
  {\bibfnamefont {L.}~\bibnamefont {Pan}}, \bibinfo {author} {\bibfnamefont
  {M.}~\bibnamefont {Lang}}, \bibinfo {author} {\bibfnamefont {Y.}~\bibnamefont
  {Jiang}}, \bibinfo {author} {\bibfnamefont {Q.}~\bibnamefont {Shao}},
  \bibinfo {author} {\bibfnamefont {T.}~\bibnamefont {Nie}}, \bibinfo {author}
  {\bibfnamefont {K.}~\bibnamefont {Murata}}, \bibinfo {author} {\bibfnamefont
  {J.}~\bibnamefont {Tang}}, \bibinfo {author} {\bibfnamefont {Y.}~\bibnamefont
  {Wang}}, \bibinfo {author} {\bibfnamefont {L.}~\bibnamefont {He}}, \bibinfo
  {author} {\bibfnamefont {T.-K.}\ \bibnamefont {Lee}}, \bibinfo {author}
  {\bibfnamefont {W.-L.}\ \bibnamefont {Lee}}, \ and\ \bibinfo {author}
  {\bibfnamefont {K.~L.}\ \bibnamefont {Wang}},\ }\href {\doibase
  10.1103/PhysRevLett.113.137201} {\bibfield  {journal} {\bibinfo  {journal}
  {Phys. Rev. Lett.}\ }\textbf {\bibinfo {volume} {113}},\ \bibinfo {pages}
  {137201} (\bibinfo {year} {2014})}\BibitemShut {NoStop}%
\bibitem [{\citenamefont {Fan}\ \emph {et~al.}(2014)\citenamefont {Fan},
  \citenamefont {Upadhyaya}, \citenamefont {Kou}, \citenamefont {Lang},
  \citenamefont {Takei}, \citenamefont {Wang}, \citenamefont {Tang},
  \citenamefont {He}, \citenamefont {Chang}, \citenamefont {Montazeri},
  \citenamefont {Yu}, \citenamefont {Jiang}, \citenamefont {Nie}, \citenamefont
  {Schwartz}, \citenamefont {Tserkovnyak},\ and\ \citenamefont
  {Wang}}]{Fan2014}%
  \BibitemOpen
  \bibfield  {author} {\bibinfo {author} {\bibfnamefont {Y.}~\bibnamefont
  {Fan}}, \bibinfo {author} {\bibfnamefont {P.}~\bibnamefont {Upadhyaya}},
  \bibinfo {author} {\bibfnamefont {X.}~\bibnamefont {Kou}}, \bibinfo {author}
  {\bibfnamefont {M.}~\bibnamefont {Lang}}, \bibinfo {author} {\bibfnamefont
  {S.}~\bibnamefont {Takei}}, \bibinfo {author} {\bibfnamefont
  {Z.}~\bibnamefont {Wang}}, \bibinfo {author} {\bibfnamefont {J.}~\bibnamefont
  {Tang}}, \bibinfo {author} {\bibfnamefont {L.}~\bibnamefont {He}}, \bibinfo
  {author} {\bibfnamefont {L.-T.}\ \bibnamefont {Chang}}, \bibinfo {author}
  {\bibfnamefont {M.}~\bibnamefont {Montazeri}}, \bibinfo {author}
  {\bibfnamefont {G.}~\bibnamefont {Yu}}, \bibinfo {author} {\bibfnamefont
  {W.}~\bibnamefont {Jiang}}, \bibinfo {author} {\bibfnamefont
  {T.}~\bibnamefont {Nie}}, \bibinfo {author} {\bibfnamefont {R.~N.}\
  \bibnamefont {Schwartz}}, \bibinfo {author} {\bibfnamefont {Y.}~\bibnamefont
  {Tserkovnyak}}, \ and\ \bibinfo {author} {\bibfnamefont {K.~L.}\ \bibnamefont
  {Wang}},\ }\href {\doibase 10.1038/nmat3973} {\bibfield  {journal} {\bibinfo
  {journal} {Nat Mater}\ }\textbf {\bibinfo {volume} {13}},\ \bibinfo {pages}
  {699} (\bibinfo {year} {2014})}\BibitemShut {NoStop}%
\bibitem [{\citenamefont {Emori}\ \emph {et~al.}(2013)\citenamefont {Emori},
  \citenamefont {Bauer}, \citenamefont {Martinez},\ and\ \citenamefont
  {Beach}}]{emorinm13}%
  \BibitemOpen
  \bibfield  {author} {\bibinfo {author} {\bibfnamefont {S.}~\bibnamefont
  {Emori}}, \bibinfo {author} {\bibfnamefont {S.-M.}\ \bibnamefont {Bauer},
  \bibfnamefont {U.and~Ahn}}, \bibinfo {author} {\bibfnamefont
  {E.}~\bibnamefont {Martinez}}, \ and\ \bibinfo {author} {\bibfnamefont
  {G.}~\bibnamefont {Beach}},\ }\href@noop {} {\bibfield  {journal} {\bibinfo
  {journal} {Nat. Mat.}\ }\textbf {\bibinfo {volume} {12}},\ \bibinfo {pages}
  {611} (\bibinfo {year} {2013})}\BibitemShut {NoStop}%
\bibitem [{\citenamefont {Ryu}\ \emph {et~al.}(2013)\citenamefont {Ryu},
  \citenamefont {Thomas}, \citenamefont {Yang},\ and\ \citenamefont
  {Parkin}}]{ryunnt13}%
  \BibitemOpen
  \bibfield  {author} {\bibinfo {author} {\bibfnamefont {K.-S.}\ \bibnamefont
  {Ryu}}, \bibinfo {author} {\bibfnamefont {L.}~\bibnamefont {Thomas}},
  \bibinfo {author} {\bibfnamefont {S.-H.}\ \bibnamefont {Yang}}, \ and\
  \bibinfo {author} {\bibfnamefont {S.}~\bibnamefont {Parkin}},\ }\href@noop {}
  {\bibfield  {journal} {\bibinfo  {journal} {Nat. Nanotech.}\ }\textbf
  {\bibinfo {volume} {8}},\ \bibinfo {pages} {527} (\bibinfo {year}
  {2013})}\BibitemShut {NoStop}%
\bibitem [{\citenamefont {Tserkovnyak}\ and\ \citenamefont
  {Bender}(2014)}]{Sh_pheno}%
  \BibitemOpen
  \bibfield  {author} {\bibinfo {author} {\bibfnamefont {Y.}~\bibnamefont
  {Tserkovnyak}}\ and\ \bibinfo {author} {\bibfnamefont {S.~A.}\ \bibnamefont
  {Bender}},\ }\href {\doibase 10.1103/PhysRevB.90.014428} {\bibfield
  {journal} {\bibinfo  {journal} {Phys. Rev. B}\ }\textbf {\bibinfo {volume}
  {90}},\ \bibinfo {pages} {014428} (\bibinfo {year} {2014})}\BibitemShut {NoStop}%
\bibitem [{\citenamefont {Hoffmann}(2013)}]{hoff_she}%
  \BibitemOpen
  \bibfield  {author} {\bibinfo {author} {\bibfnamefont {A.}~\bibnamefont
  {Hoffmann}},\ }\href {\doibase 10.1109/TMAG.2013.2262947} {\bibfield
  {journal} {\bibinfo  {journal} {Magnetics, IEEE Transactions on}\ }\textbf
  {\bibinfo {volume} {49}},\ \bibinfo {pages} {5172} (\bibinfo {year}
  {2013})}\BibitemShut {NoStop}%
\bibitem [{\citenamefont {Dzyaloshinskii}(1957)}]{Dzioloshinskii}%
  \BibitemOpen
  \bibfield  {author} {\bibinfo {author} {\bibfnamefont {I.~E.}\ \bibnamefont
  {Dzyaloshinskii}},\ }\href@noop {} {\bibfield  {journal} {\bibinfo  {journal}
  {Sov. Phys. JETP}\ }\textbf {\bibinfo {volume} {5}},\ \bibinfo {pages} {1259}
  (\bibinfo {year} {1957})}\BibitemShut {NoStop}%
\bibitem [{\citenamefont {Moriya}(1960)}]{Moriya}%
  \BibitemOpen
  \bibfield  {author} {\bibinfo {author} {\bibfnamefont {T.}~\bibnamefont
  {Moriya}},\ }\href {\doibase 10.1103/PhysRev.120.91} {\bibfield  {journal}
  {\bibinfo  {journal} {Phys. Rev.}\ }\textbf {\bibinfo {volume} {120}},\
  \bibinfo {pages} {91} (\bibinfo {year} {1960})}\BibitemShut {NoStop}%
\bibitem [{\citenamefont {Thiele}(1973)}]{Thiele}%
  \BibitemOpen
  \bibfield  {author} {\bibinfo {author} {\bibfnamefont {A.~A.}\ \bibnamefont
  {Thiele}},\ }\href {\doibase 10.1103/PhysRevLett.30.230} {\bibfield
  {journal} {\bibinfo  {journal} {Phys. Rev. Lett.}\ }\textbf {\bibinfo
  {volume} {30}},\ \bibinfo {pages} {230} (\bibinfo {year} {1973})}\BibitemShut
  {NoStop}%
\bibitem [{\citenamefont {Tretiakov}\ \emph {et~al.}(2008)\citenamefont
  {Tretiakov}, \citenamefont {Clarke}, \citenamefont {Chern}, \citenamefont
  {Bazaliy},\ and\ \citenamefont {Tchernyshyov}}]{Bazaliy2008}%
  \BibitemOpen
  \bibfield  {author} {\bibinfo {author} {\bibfnamefont {O.~A.}\ \bibnamefont
  {Tretiakov}}, \bibinfo {author} {\bibfnamefont {D.}~\bibnamefont {Clarke}},
  \bibinfo {author} {\bibfnamefont {G.-W.}\ \bibnamefont {Chern}}, \bibinfo
  {author} {\bibfnamefont {Y.~B.}\ \bibnamefont {Bazaliy}}, \ and\ \bibinfo
  {author} {\bibfnamefont {O.}~\bibnamefont {Tchernyshyov}},\ }\href {\doibase
  10.1103/PhysRevLett.100.127204} {\bibfield  {journal} {\bibinfo  {journal}
  {Phys. Rev. Lett.}\ }\textbf {\bibinfo {volume} {100}},\ \bibinfo {pages}
  {127204} (\bibinfo {year} {2008})}\BibitemShut {NoStop}%
\bibitem [{\citenamefont {Schryer}\ and\ \citenamefont
  {Walker}(1974)}]{walker1974}%
  \BibitemOpen
  \bibfield  {author} {\bibinfo {author} {\bibfnamefont {N.~L.}\ \bibnamefont
  {Schryer}}\ and\ \bibinfo {author} {\bibfnamefont {L.~R.}\ \bibnamefont
  {Walker}},\ }\href {\doibase 10.1063/1.1663252} {\bibfield  {journal}
  {\bibinfo  {journal} {J. Appl. Phys.}\ }\textbf {\bibinfo {volume} {45}},\
  \bibinfo {pages} {5406} (\bibinfo {year} {1974})}\BibitemShut {NoStop}%
\bibitem [{\citenamefont {{Landau}}\ and\ \citenamefont
  {{Lifshitz}}(1980)}]{landauBKOns}%
  \BibitemOpen
  \bibfield  {author} {\bibinfo {author} {\bibfnamefont {L.}~\bibnamefont
  {{Landau}}}\ and\ \bibinfo {author} {\bibfnamefont {E.}~\bibnamefont
  {{Lifshitz}}},\ }\href@noop {} {\emph {\bibinfo {title} {Statistical Physics,
  Part 1: Volume 5}}}\ (\bibinfo  {publisher} {Pergamon, Oxford},\ \bibinfo
  {year} {1980})\BibitemShut {NoStop}%
\bibitem [{\citenamefont {{Pitaevskii}}\ and\ \citenamefont
  {{Lifshitz}}(1980)}]{landauBK80}%
  \BibitemOpen
  \bibfield  {author} {\bibinfo {author} {\bibfnamefont {L.}~\bibnamefont
  {{Pitaevskii}}}\ and\ \bibinfo {author} {\bibfnamefont {E.}~\bibnamefont
  {{Lifshitz}}},\ }\href@noop {} {\emph {\bibinfo {title} {Statistical Physics,
  Part 2: Volume 9}}}\ (\bibinfo  {publisher} {Butterwoth-Heinemann},\ \bibinfo
  {year} {1980})\BibitemShut {NoStop}%
\bibitem [{\citenamefont {Thiaville}\ \emph {et~al.}(2012)\citenamefont
  {Thiaville}, \citenamefont {Rohart}, \citenamefont {Jue}, \citenamefont
  {Cros},\ and\ \citenamefont {Fert}}]{Thiaville12}%
  \BibitemOpen
  \bibfield  {author} {\bibinfo {author} {\bibfnamefont {A.}~\bibnamefont
  {Thiaville}}, \bibinfo {author} {\bibfnamefont {S.}~\bibnamefont {Rohart}},
  \bibinfo {author} {\bibfnamefont {E.}~\bibnamefont {Jue}}, \bibinfo {author}
  {\bibfnamefont {V.}~\bibnamefont {Cros}}, \ and\ \bibinfo {author}
  {\bibfnamefont {A.}~\bibnamefont {Fert}},\ }\href@noop {} {\bibfield
  {journal} {\bibinfo  {journal} {Europhys. Lett.}\ }\textbf {\bibinfo {volume}
  {100}},\ \bibinfo {pages} {57002} (\bibinfo {year} {2012})}\BibitemShut
  {NoStop}%
\bibitem [{\citenamefont {B\"{u}ttiker}\ \emph {et~al.}(1994)\citenamefont
  {B\"{u}ttiker}, \citenamefont {Thomas},\ and\ \citenamefont
  {Pr\^etre}}]{Buttikerpump}%
  \BibitemOpen
  \bibfield  {author} {\bibinfo {author} {\bibfnamefont {M.}~\bibnamefont
  {B\"{u}ttiker}}, \bibinfo {author} {\bibfnamefont {H.}~\bibnamefont
  {Thomas}}, \ and\ \bibinfo {author} {\bibfnamefont {A.}~\bibnamefont
  {Pr\^etre}},\ }\href {\doibase 10.1007/BF01307664} {\bibfield  {journal}
  {\bibinfo  {journal} {Z. Phys. B}\ }\textbf {\bibinfo {volume} {94}},\
  \bibinfo {pages} {133} (\bibinfo {year} {1994})}\BibitemShut {NoStop}%
\bibitem [{\citenamefont {Brouwer}(1998)}]{brouwer98}%
  \BibitemOpen
  \bibfield  {author} {\bibinfo {author} {\bibfnamefont {P.~W.}\ \bibnamefont
  {Brouwer}},\ }\href {\doibase 10.1103/PhysRevB.58.R10135} {\bibfield
  {journal} {\bibinfo  {journal} {Phys. Rev. B}\ }\textbf {\bibinfo {volume}
  {58}},\ \bibinfo {pages} {R10135} (\bibinfo {year} {1998})}\BibitemShut
  {NoStop}%
\bibitem [{Note1()}]{Note1}%
  \BibitemOpen
  \bibinfo {note} {Although, in principle, the fixed external magnetic field
  breaks the time-reversal symmetry, we are interested in the regime where the
  associated Zeeman energy scale is much smaller than the exchange-correlation
  energy. In this case, the dispersion relations for chiral modes are not
  significantly affected by the external field}\BibitemShut {NoStop}%
\bibitem [{\citenamefont {{Takei}}\ \emph {et~al.}(2015)\citenamefont
  {{Takei}}, \citenamefont {{Yacoby}}, \citenamefont {{Halperin}},\ and\
  \citenamefont {{Tserkovnyak}}}]{So_graphene}%
  \BibitemOpen
  \bibfield  {author} {\bibinfo {author} {\bibfnamefont {S.}~\bibnamefont
  {{Takei}}}, \bibinfo {author} {\bibfnamefont {A.}~\bibnamefont {{Yacoby}}},
  \bibinfo {author} {\bibfnamefont {B.~I.}\ \bibnamefont {{Halperin}}}, \ and\
  \bibinfo {author} {\bibfnamefont {Y.}~\bibnamefont {{Tserkovnyak}}},\
  }\href@noop {} {\bibfield  {journal} {\bibinfo  {journal} {ArXiv e-prints}\ }
  (\bibinfo {year} {2015})},\ \Eprint {http://arxiv.org/abs/1506.01061}
  {arXiv:1506.01061 [cond-mat.mes-hall]} \BibitemShut {NoStop}%
\bibitem [{\citenamefont {Zhang}\ \emph {et~al.}(2009)\citenamefont {Zhang},
  \citenamefont {Liu}, \citenamefont {Qi}, \citenamefont {Dai}, \citenamefont
  {Fang},\ and\ \citenamefont {Zhang}}]{Zhang2009}%
  \BibitemOpen
  \bibfield  {author} {\bibinfo {author} {\bibfnamefont {H.}~\bibnamefont
  {Zhang}}, \bibinfo {author} {\bibfnamefont {C.-X.}\ \bibnamefont {Liu}},
  \bibinfo {author} {\bibfnamefont {X.-L.}\ \bibnamefont {Qi}}, \bibinfo
  {author} {\bibfnamefont {X.}~\bibnamefont {Dai}}, \bibinfo {author}
  {\bibfnamefont {Z.}~\bibnamefont {Fang}}, \ and\ \bibinfo {author}
  {\bibfnamefont {S.-C.}\ \bibnamefont {Zhang}},\ }\href {\doibase
  10.1038/nphys1270} {\bibfield  {journal} {\bibinfo  {journal} {Nature
  Physics}\ }\textbf {\bibinfo {volume} {5}},\ \bibinfo {pages} {438} (\bibinfo
  {year} {2009})}\BibitemShut {NoStop}%
\bibitem [{\citenamefont {Kou}\ \emph {et~al.}(2012)\citenamefont {Kou},
  \citenamefont {Jiang}, \citenamefont {Lang}, \citenamefont {Xiu},
  \citenamefont {He}, \citenamefont {Wang}, \citenamefont {Wang}, \citenamefont
  {Yu}, \citenamefont {Fedorov}, \citenamefont {Zhang},\ and\ \citenamefont
  {Wang}}]{Xufeng2012}%
  \BibitemOpen
  \bibfield  {author} {\bibinfo {author} {\bibfnamefont {X.~F.}\ \bibnamefont
  {Kou}}, \bibinfo {author} {\bibfnamefont {W.~J.}\ \bibnamefont {Jiang}},
  \bibinfo {author} {\bibfnamefont {M.~R.}\ \bibnamefont {Lang}}, \bibinfo
  {author} {\bibfnamefont {F.~X.}\ \bibnamefont {Xiu}}, \bibinfo {author}
  {\bibfnamefont {L.}~\bibnamefont {He}}, \bibinfo {author} {\bibfnamefont
  {Y.}~\bibnamefont {Wang}}, \bibinfo {author} {\bibfnamefont {Y.}~\bibnamefont
  {Wang}}, \bibinfo {author} {\bibfnamefont {X.~X.}\ \bibnamefont {Yu}},
  \bibinfo {author} {\bibfnamefont {A.~V.}\ \bibnamefont {Fedorov}}, \bibinfo
  {author} {\bibfnamefont {P.}~\bibnamefont {Zhang}}, \ and\ \bibinfo {author}
  {\bibfnamefont {K.~L.}\ \bibnamefont {Wang}},\ }\href {\doibase
  http://dx.doi.org/10.1063/1.4754452} {\bibfield  {journal} {\bibinfo
  {journal} {Journal of Applied Physics}\ }\textbf {\bibinfo {volume} {112}},\
  \bibinfo {eid} {063912} (\bibinfo {year} {2012})}\BibitemShut {NoStop}%
\bibitem [{Note2()}]{Note2}%
  \BibitemOpen
  \bibinfo {note} {The underlying adiabatic approximation, used in deriving the
  pumping, is valid for values of $v$ such that the effective bias in domain
  wall's reference frame $V_{\protect \rm eff}$ (as discussed in the main text)
  is much less than the gap $2\Delta $. Substituting for $V_{\protect \rm eff}$
  from the main text this becomes equivalent to $v \ll v_F$, which is satisfied
  for the chosen material parameters.}\BibitemShut {Stop}%
\end{thebibliography}
\end{document}